\documentclass[%
reprint,
superscriptaddress,
 amsmath,amssymb,
 aps,
pra,
floatfix,
]{revtex4-1}

\usepackage{graphicx}
\usepackage{dcolumn}
\usepackage{bm}
\usepackage{xcolor}
\usepackage{soul}

\begin{document}
\setstcolor{red}
\preprint{APS/123-QED}

\title{Shaping THz emission spectra by using \\ sub-wavelength nanopatterned spintronic THz emitters}

\author{Bikash Das-Mohapatra}
\affiliation{Institut f\"{u}r Physik, Martin-Luther-Universit\"{a}t Halle-Wittenberg, Von-Danckelmann-Platz 3, 06120 Halle, Germany}

\author{Reza Rouzegar}
\affiliation{Department of Physical Chemistry, Fritz Haber Institute, Faradayweg 4-6, 14195 Berlin, Germany}

\author{Evangelos Th. Papaioannou}
\affiliation{Institut f\"{u}r Physik, Martin-Luther-Universit\"{a}t Halle-Wittenberg, Von-Danckelmann-Platz 3, 06120 Halle, Germany}
\affiliation{Department of Physics, Aristotle University of Thessaloniki, 54124 Thessaloniki, Greece}
 
\author{Tobias Kampfrath}
\affiliation{Department of Physical Chemistry, Fritz Haber Institute, Faradayweg 4-6, 14195 Berlin, Germany}

\author{Georg Schmidt}
\email{georg.schmidt@physik.uni-halle.de}
\affiliation{Institut f\"{u}r Physik, Martin-Luther-Universit\"{a}t Halle-Wittenberg, Von-Danckelmann-Platz 3, 06120 Halle, Germany}
\affiliation{Interdisziplin\"{a}res Zentrum f\"{u}r Materialwissenschaften,
Martin-Luther-Universit\"{a}t Halle-Wittenberg, Heinrich-Damerow-Strasse 4, 06120 Halle, Germany
}

\date{\today}

\begin{abstract}
We show in theory and experiment that in periodically patterned spintronic THz emitters (STE),  charge dynamics can modify the emission spectrum in a well-controlled way. Characterization of sub-wavelength patterned STE at frequencies up to 30\,THz shows that the STE's emission spectrum systematically changes with emitter size. The spectral intensity exhibits significant reductions at frequencies below 4\,THz, accompanied by pronounced dips at around 15\,THz and 24\,THz. While reducing the STE size enhances the modulation of all features, it does not alter the dip frequencies. The effect originates from the charging of the structure's edges by THz currents, causing a backflow that interferes with the primary current pulse. An analytical model quantitatively reproduces these results and agrees well with control experiments. Our findings enable a detailed investigation of the charge dynamics in STE and provide additional means for controlled shaping of STE emission spectra by nano patterning.
\end{abstract}

\maketitle


\section{Introduction}

Spintronic THz emitters (STEs) present a new route to creating broadband THz radiation. STEs usually consist of a ferromagnet (FM)/non-magnet (NM) metallic heterostructure\cite{Seifert2016,Wu_Elyasi_ADMA,Kimel_Nat_Nano_2016} but also many other thin film material systems have been investigated\cite{Papaioannou2021,Jaffres2020,Thomas2021,yigPt2018,Albrecht2019,Albrecht2020,2D-HMH_Nat_Comm}. 
The properties of the emitted THz signal have been studied with respect to: thicknesses of the layers\cite{Seifert2016,Torosyan2018,Qiu:18,Wu_Elyasi_ADMA}, geometrical stacking order\cite{ADOM:ADOM201600270,Seifert2016,Haifeng2018,Ogasawara_2020,Li_2019_Japan,Chen_circular}, quality of interfaces\cite{Sasaki2017,Li_2019,Papa2019} and interface materials\cite{Oliver2021,Scheuer_2022,Li_2018,Jacques2021,rouzegar2023}, and excitation wavelength\cite{Papaioannou2018,Herapath2019}. Further, the engineering of time and frequency profile of the THz pulse has been investigated by applying different antenna-coupled designs to the emitters.\cite{elezzabi_apl21,Talara_2021,Nandi2019}
In all these devices, an ultrafast laser pulse induces an ultra-short spin-current pulse\cite{rouzegar2022} from the ferromagnet into the non-magnet which therein is converted into a lateral charge current pulse via the inverse spin-Hall effect (ISHE)\cite{Kampfrath2013}. 

Depending on the application of the STE in THz spectroscopy or THz photonics, it is highly desirable to tune the spectrum of the emitted THz pulse. Recently, it has been shown theoretically that local charge accumulation and a subsequent relaxation process can strongly modify the emission spectrum, especially by suppressing low frequency components \cite{chargedynamics}. It seemed, however, difficult to demonstrate this effect in the experiment. On the one hand, it is difficult to extract the relevant geometrical parameters in terms of conductance and capacitance from the layer structure that are necessary for the exact analysis. On the other hand, modifying the layer structure of the STE can influence the emission characteristics in more than one way \cite{Gorchon22}, making it almost impossible to relate modifications of the spectrum directly to these charge dynamics. 

We have solved this problem by designing a set of experiments using arrays of nanopatterned STE (namely squares and rectangles) that allow us to demonstrate and control these charge dynamics. With a size of down to less than 1 micrometer, the squares and rectangles under investigation are smaller than the exciting laser spot (30\,$\mathrm{\mu}$m), ensuring a homogeneous illumination and thus creating boundary conditions different from those for a limited laser spot on a large area emitter. Also, both the size and period of the arrays are smaller than the wavelength at 100\,THz in vacuum to avoid interference effects in the observed spectral range.

In recent past, a number of groups have investigated the influence of patterning on the THz emission of STE. For example, the emission from 5\,$\mathrm{\mu m}$ wide stripe patterns \cite{doi:10.1002/pssr.201900057,ADOM:ADOM201600270} showed reduced intensity and a blue shift. Other experiments with stripes of several hundred micron width show no frequency shift of the intensity maximum (1\,THz) but an increase at lower frequencies that the authors attribute to multi slit interference \cite{Wu2020}. Finally, Song et al. \cite{Song_2019} investigated rectangles with a short axis varying from 20\,$\mu$m to 320\,$\mu$m. They also observed a strongly decreased intensity and an apparent blue shift.
It should be noted that none of these experiments investigated the spectral range above 5\,THz.

In our experiments, we use structures more than one order of magnitude smaller than those investigated before \cite{ADOM:ADOM201600270, doi:10.1002/pssr.201900057,          Wu2020, Song_2019}. We also detect a spectral range up to frequencies as high as 30\,THz which gives us access to the whole range of frequencies that are influenced by the charge dynamics.

\section{Results and Discussion}
\subsection{Sample Fabrication}
\begin{figure}

{\includegraphics{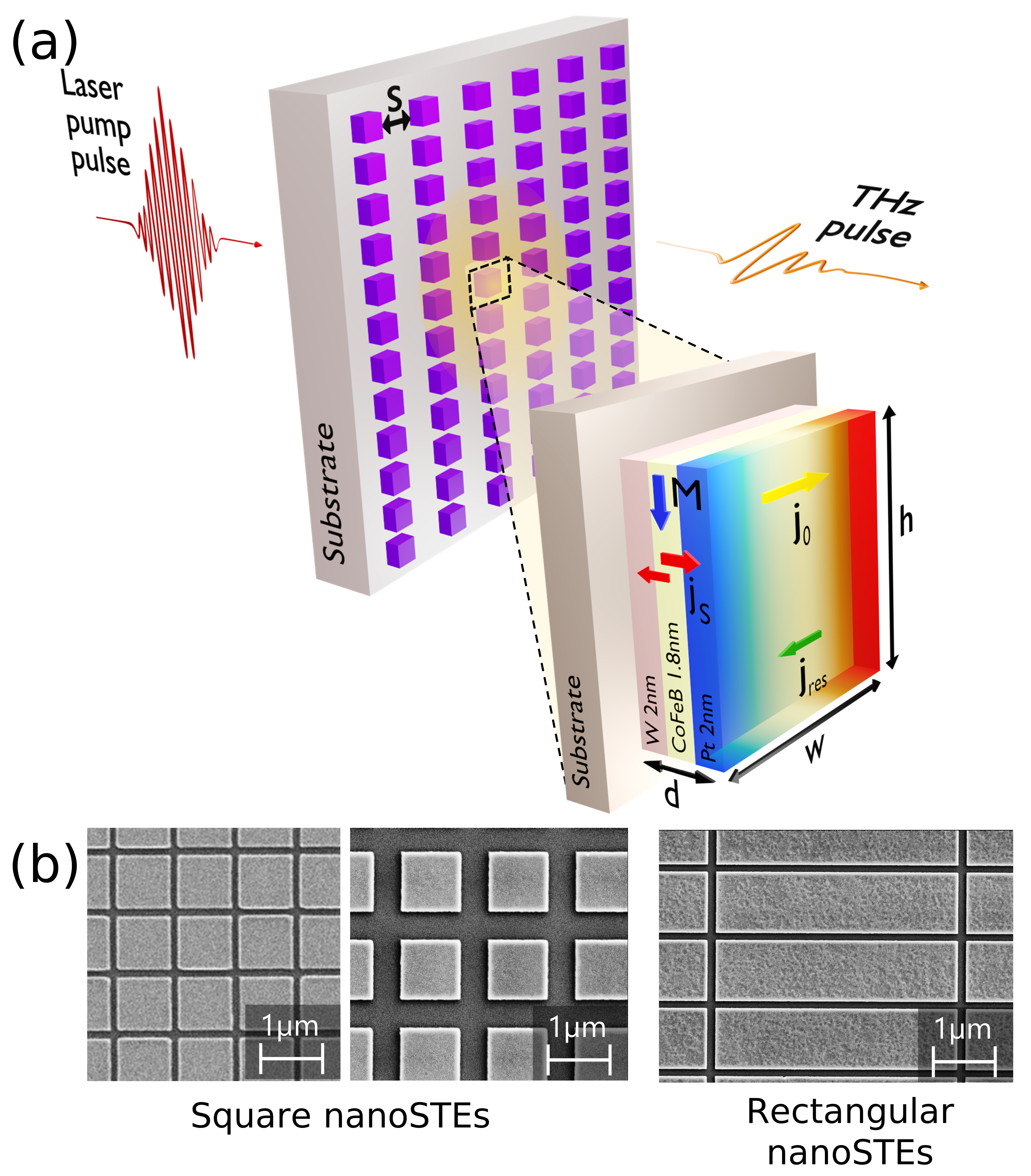}}
   \caption{(a) Sketch of THz emission by optically excited nanopatterned STEs. The pattern is a two-dimensional lattice of rectangles of widths($w$), height($h$), thickness($d$) and spacing($s$). Here we use a trilayer STE [W(2\,nm)/CoFeB(1.8\,nm)/Pt(2\,nm)]. (b) SEM images of various STEs nanopatterned into square and rectangular nanostructures with different $w$, $h$ and $s$.}
    \label{fig1}
  \end{figure}

We use a trilayer Spintronic Terahertz Emitter (STE) with the layer stacking W(2\,nm) / $\mathrm{Co_{20}Fe_{60}B_{20}}$(1.8\,nm) / Pt(2\,nm) (Fig. \ref{fig1}(a)). As shown by Seifert et al.\cite{Seifert2016,upscaling_2017}, the trilayer provides optimum Terahertz output. Spin currents excited by an ultrafast laser pulse, flow in opposite directions into the top (Pt) and bottom (W) layer respectively, where they are converted to parallel in plane charge current pulses due to the opposite signs of the respective spin-Hall angles. The total transient current emits a THz pulse easily covering the 1-30\,THz range when excited with 10\,fs laser pulses\cite{Seifert2016}. This metallic heterostructure is grown on Sapphire substrate using ultrahigh-vacuum DC magnetron sputtering. 
On the stack, a thin film of PMMA is spin coated. Using positive electron beam lithography the resist is patterned. After evaporation of 30\,nm Al$_2$O$_3$ and lift-off, the remaining Al$_2$O$_3$ structures serve as an etch mask for patterning by Ar-ion milling of the desired structures. With this process we fabricate square emitters with lateral sizes `$w$' of 900\,nm, 1.8\,$\mu$m, 4.5\,$\mu$m, 9.0\,$\mu$m and 18.0\,$\mu$m, respectively. To achieve a signal that allows for quantitative analysis we use arrays of squares that are much larger than the laser spot diameter of 30\,$\mu$m and that have a constant filling factor of 81\% by using a spacing `$s$' in both lateral dimensions of 1/9 of the square size. Fig. \ref{fig1}(b) shows scanning electron microscope (SEM) images of square and rectangular emitters.

\subsection{Experimental Setup}

The nanopatterned STE is excited with linearly polarized laser pulses (wavelength of 800\,nm, duration of 10\,fs and pulse energy of 2\,nJ) from a Ti:sapphire laser oscillator (repetition rate at 80 MHz) under normal incidence. The pump beam diameter at the sample position is approximately 30\,$\mu$m full width at half maximum of the intensity. The resulting transient charge current described in the previous section causes a transient THz electric field $E(t)$, whose far field is detected by electrooptic sampling where a probe pulse (0.6\,nJ, 10\,fs) copropagates with the terahertz pulse through an electrooptic crystal. As electrooptical crystal,  ZnTe(110) (thickness of 10\,$\mu$m) is used. The probe pulse's ellipticity is measured as a function of the delay between terahertz and sampling pulse by means of a balanced detection (consisting of a quarter wave-plate, a polarizing beam splitter, two balanced photodiode and a half wave-plate). All experiments are performed at room temperature in ambient conditions.

\label{sec:examples}

\subsection{THz emission from nanostructured STEs}

\begin{figure}

{\includegraphics{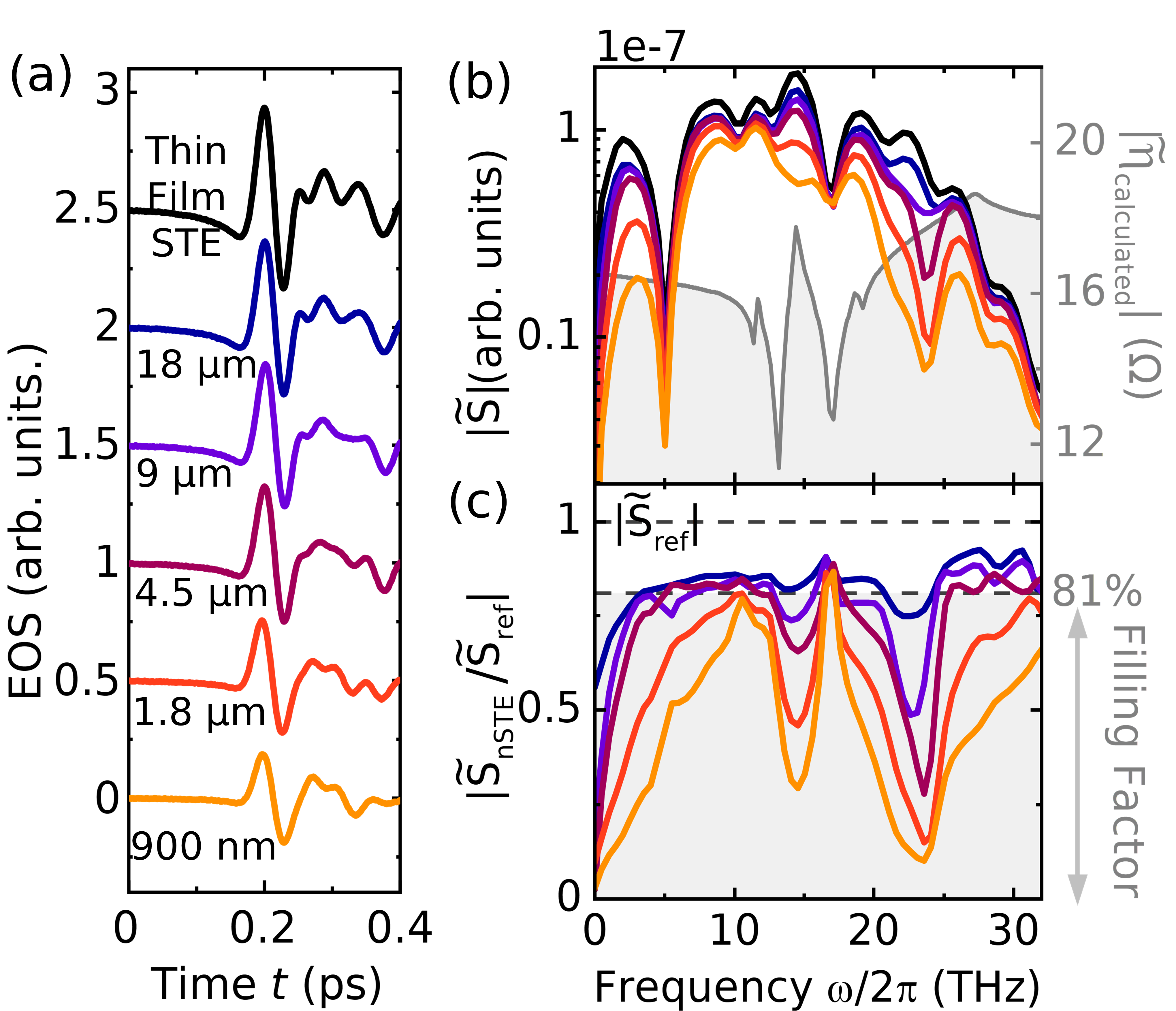}}
\caption{(a)Electro-optic signal (EOS) from patterned STEs with different square sizes with the dimensions, $w$ = $h$ = 900\,nm; 1.8$\mu$m; 4.5$\mu$m; 9$\mu$m; 18$\mu$m; all with a filling factor of 81$\%$. (b) Amplitude Spectra $\mathrm{|\tilde S(\omega)|}$ obtained for different square sizes by FFT of the electro-optic signal from (a). The grey line shows the calculated characteristic wave impedance $|\tilde{\eta}(\omega)|$ for the Sapphire substrate. (c) Ratio between spectra for nanoSTEs and thin film STE to indicate the relative change in the spectra.}
\label{fig2}
\end{figure}

In Fig. \ref{fig2}(a) we show the respective detected signals for differently sized square emitters in the time domain. The peak-to-peak amplitude that is often used as a reference, slightly decreases with decreasing structure size and for the smallest structures of 900\,nm, it is less than 50\% of the signal for a continuous STE film. 

Fig. \ref{fig2}(b) shows the spectral amplitude for the different samples. The spectra were obtained from the data in Fig. \ref{fig2}(a) by fast Fourier transform (FFT) of the data using a von Hann window \cite{Essenwanger_1986}. For an unpatterned STE, we observe emission up to $\sim$30\,THz within the dynamic range of the measurement. 

The dip at 5\,THz (Fig. \ref{fig2}(b)) arises from the zero of the electro-optic response of the ZnTe crystal\cite{Leitenstorfer1999}, the dip at 10\,THz originates from the Ge wafer behind the STE\cite{Singh2018}. Further dips\cite{Seifert2016} at 13, 17 and 19\,THz are due to the refractive index \cite{1985querry,Gueckstock:21} of the Sapphire substrate which can be also observed from the calculated STE impedance $\mathrm{|\tilde{\eta}(\omega)|}$ including the Sapphire substrate as shown in grey in Fig. \ref{fig2}(b).

Interestingly, for the patterned structures, we observe a small overall decrease in amplitude but also two distinct features that become more prominent for smaller structures. Firstly, frequencies below approx. 4\,THz are strongly suppressed by approx. a factor of five for the smallest structures compared to the large area emitter. Secondly, a decrease by a similar factor can be observed at frequencies of  approx. 15\,THz and 24\,THz, resulting two dips in the spectral amplitude. It should be noted that for higher frequencies, the amplitudes are again almost identical for all structure sizes and that the frequencies of the dips do not depend on structure size making an interference effect highly unlikely. While Fig. \ref{fig2}(b) shows the different spectral amplitude for the different respective structures, Fig. \ref{fig2}(c) shows the filter function, i.e., the ratio of the spectral amplitude for the respective nanoemitters divided by that of the large area reference emitter, (S$\mathrm{_{nSTE}(\omega)}$/S$\mathrm{_{ref}(\omega)}$). Here, the relative change is even more pronounced.  

\subsection{Charge Accumulation in nanostructured STEs}
To understand the filter function of Fig. \ref{fig2}(c) we have to understand how geometry controls the charge dynamics. In a large area emitter, the laser pulse initially creates a lateral current density which is proportional to the local light intensity. This leads to local charging at the edges of the spot resulting in a backflow inside and outside the spot as described in \cite{chargedynamics}. The patterned STE has the additional boundary condition that, outside the illuminated area (namely outside the square), we have $j=0$ limiting the backflow to inside the illuminated area. With uniform illumination charging occurs at opposite edges of the structure where $\vec\nabla\cdot\vec j\ne 0$ and thus $\partial\rho/\partial t\neq0$. Because we have a homogeneous current density perpendicular to the edges, we also get homogeneous charging along the edges and a resulting backflow perpendicular to the edges and against the initial current direction. So we are now dealing with a quasi one dimensional problem. Note that the charged area is very narrow and only extends over the Thomas-Fermi screening length which is negligible compared to the total width of the structures $w$. 

\begin{figure}
{\includegraphics{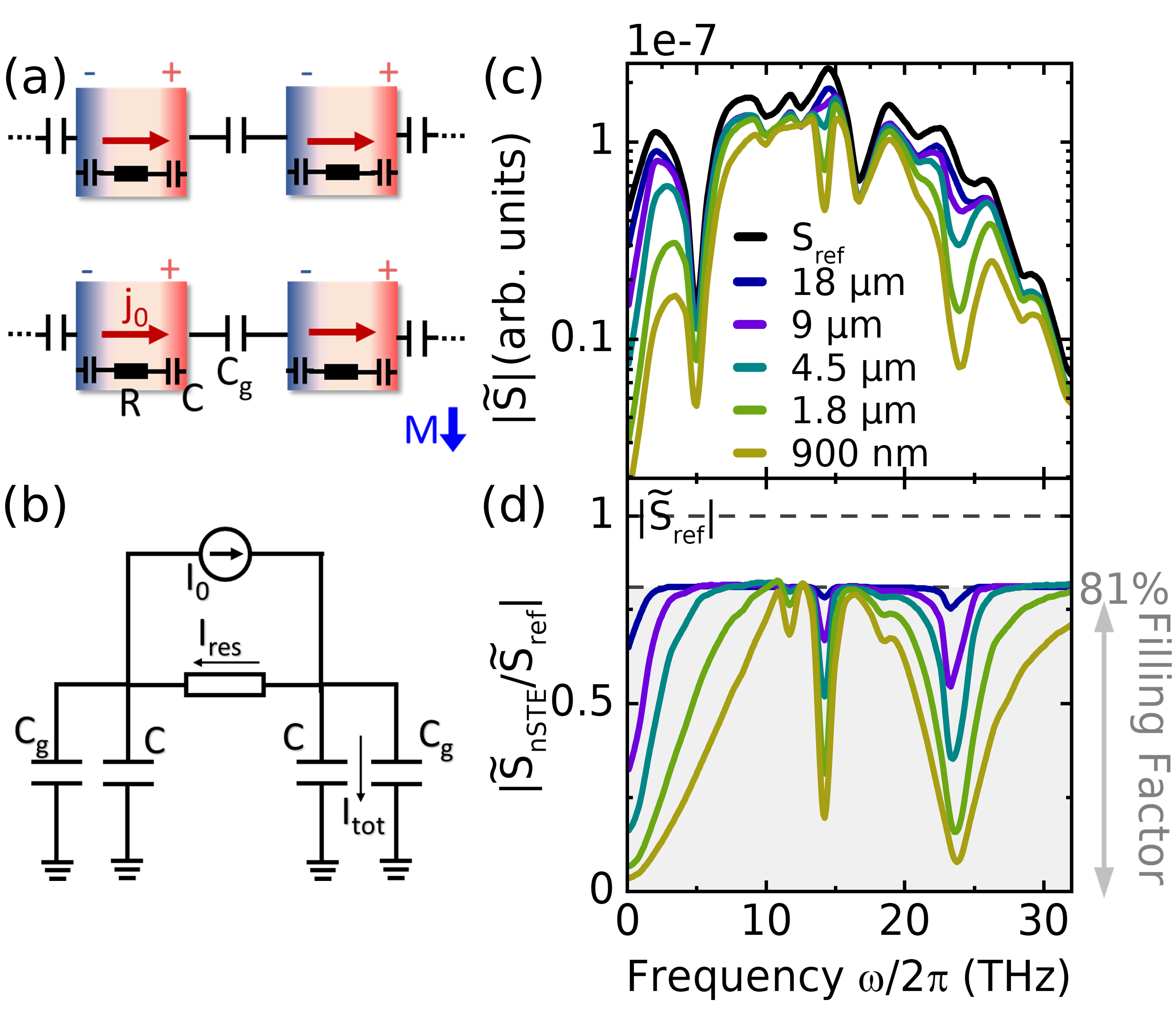}}
\caption{(a) Sketch of nanoSTEs showing the charge current density $\vec j_0$, the edge cpacitance $C$, the capacitance between emitters $C_g$, the resistance in the nanoSTE $\mathrm{R}$, and the charge accumulation along the edges. (b)Equivalent R-C circuit for the nanoSTE unit cell. (c)Spectra  $\mathrm{|\tilde S(\omega)|}$ for different sized nanoSTEs calculated by applying the circuit Model on the spectrum of the thin film reference STE. (d)Ratio or transfer function of the same spectra  showing good agreement with the experimental data in Fig. \ref{fig2}(c). }
\label{fig3}
\end{figure}

For an analytical model, we need to determine the relevant time constants. Two edges of the structure along $h$ can be seen as capacitors $C$ that are charged with opposite charges. Opposite edges do not act as two plates of a single capacitor but are independent capacitors as shown in Fig. \ref{fig3}(a). If a second structure is in the vicinity, an additional capacitance can appear due to the coupling between structures now creating two opposite plates of a plate capacitor labeled $C_g$. The opposite charges from the two opposite edges equalize by a current through the resistance $R$ which is given by the total resistance along the x-axis that connects the two edge capacitors. Although it is difficult to calculate the absolute values of $C$, $C_g$, and $R$, we can make the following assumptions.

$C$ is determined by the geometry and by the complex dielectric functions $\epsilon$ of the STE and the substrate. Because $\epsilon$ is frequency dependent, $C$ also depends on frequency. In addition, we know that, as long as $w$ is larger than the relevant screening length (which is always the case in our experiment), $C$ is independent from $w$. Moreover, $C$ scales linearly with $h$. $C_g$ is also independent of $w$ and proportional to $h$. However, it also should be roughly inversely proportional to the spacing $s$. The resistance R is given by the resistivity in x-direction, inversely proportional to $h$ and proportional to the $w$. As a result, we find that changing $w$ changes only the resistance (but not the capacitance) so that RC increases linearly with $w$. Changing $h$ changes $R$, $C$ and $C_g$ but keeps $R(C+C_g)$ (which we discuss as $RC$ in the following) constant. Note that while RC remains constant, the total backflow current is proportional to $h$.

As a consequence changing $w$ changes the relevant time constant while changing $h$ does not.    

\subsection{Modelling the THz spectrum}
As justified above, we assume a one dimensional current flow, which is valid except for small corrections and a time dependent current density $\vec j(t,x)$ that only depends on x but is independent of the y and (because the metal stack is only a few nm thick) also of the z coordinate. This allows us to use a current $\vec I(t,x)=\vec j(t,x)\times h \times d$. Because current density and current only have an x-component we do without the vector sign and use $j(t,x)$ and $I(t,x)$.
The laser illumination creates a homogeneous current density (also independent of x) in the emitter which we label $j_0(t)$ and which yields a current $I_0(t)=j_0(t)\times h \times d$. This current results in a far field emission proportional to $I_0(t)\cdot w$. On the other hand the current $I_0$ also charges the capacitances $C$ and $C_g$. Discharge of these capacitors happens via the response current $\vec I_{res}(t,x)=\vec j_{res}(t,x)\cdot h \cdot d$, that depends on the potential that builds up due to the charges at the edges. But again the negative charge at one end needs to flow through the whole structure and thus through the resistance $R$ to compensate the positive charge at the other end (Fig. \ref{fig3}(a)). 

For the modelling we can use the equivalent circuit of Fig. \ref{fig3}(b). $I_0$ charges $C$ and $C_g$ and the discharge happens via R. Because we have $C$ and $C_g$ in parallel we can simplify further by only dealing with $C$ and later adding the contribution of $C_g$. The equivalent circuit is then described by the following differential equation:
\begin{equation}
    C\frac{dU_C(t)}{dt}=I_{0}(t)-2\cdot\frac{U_C(t)}{R}
\end{equation}

where U$_C$ is the voltage on the capacitor and -U$_C$/R corresponds to $I_{res}$. We can do a Fourier transform of this differential equation that yields

\begin{equation}
    i\omega C\tilde{U}_C(\omega)=\tilde{I}_{0}(\omega)-2\cdot\frac{\tilde{U}_C(\omega)}{R}
\end{equation}

With

\begin{equation}
    \tilde{I}_{tot}(\omega)=\tilde{I}_{0}(\omega)+\tilde{I}_{res}(\omega)
\end{equation}

this yields

\begin{equation}
    \tilde{I}_{tot}(\omega)=i\omega C\tilde{U}_C(\omega) =\tilde{I}_{0}(\omega)\frac{1}{1+\frac{2}{i\omega RC}}=\tilde F_T(\omega)\tilde{I}_{0}(\omega)
\end{equation}

$\tilde F_T(\omega)$ is the transfer function by which the current $\tilde{I}_{0}(\omega)$ in a large area reference emitter needs to be multiplied to get the response of the nanoSTE. This shows that $\tilde{I}_{tot}(\omega)$ for the nanoSTE is mainly the initial current $\tilde{I}_{0}(\omega)$ modified by a high pass filter function. While in \cite{chargedynamics} the $RC$ constant of the high pass filter only depended on the material parameters, we now have $C$ and $R$ geometry dependent and we can control $RC$ by changing the width of the structure $w$. For the large area reference with $RC\rightarrow\infty$, this yields $\tilde{I}_0$ or a transfer function $F_T$ which is 1 independent from frequency, as expected. It should be noted that our model assumes that the $RC$ constant of the large area emitter is large enough to be neglected in comparison to that of the nanoSTE. In the following we assume that the spectrum of the reference STE is given by $I_0$.

\begin{figure*}
\centering
{\includegraphics[width=0.85\textwidth]{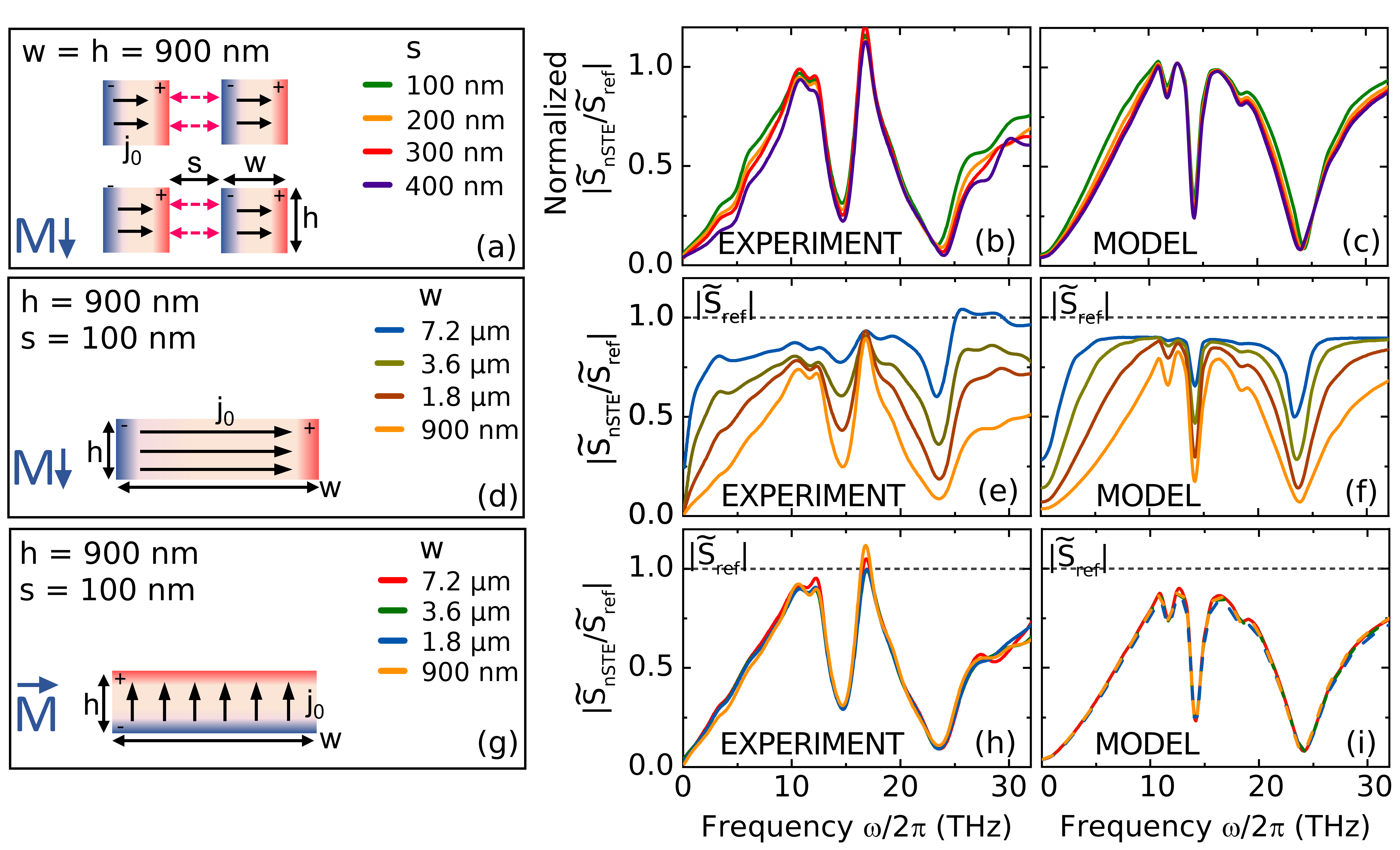}}
\caption{(a)$\mathrm{|\tilde S_{nSTE}(\omega)/\tilde S_{ref}(\omega)|}$ (normalized based on filling factor) obtained for an array of squares of size 900\,nm and different spacing, $s$ = 100\,nm, 200\,nm, 400\,nm and 800\,nm from (b)the experiment and (c)the model. (d)$\mathrm{|\tilde S_{nSTE}(\omega)/\tilde S_{ref}(\omega)|}$ for an array of rectangles of $h$ = 900\,nm and $s$ = 100\,nm and different width, $w$ = 900\,nm, 1.8\,µm, 3.6\,µm and 7.2\,µm when the magnetization is perpendicular to the width ((e) experiment and (f) model) and (g)parallel to the width ((h) experiment and (i) model). The sketches in (a), (d) and (g) show the magnetization $\vec M$ and the corresponding charge current $\vec j_0$ direction in the respective nanoSTE. The + and - signs represent the charge accumulation.}
\label{fig4}
\end{figure*}

The time constant RC can be expressed as $k\cdot\epsilon(\omega)/\sigma(\omega)$ where the factor $k$ takes into account the geometry of the nanoSTE and $\epsilon(\omega)$ and $\sigma(\omega)$ are the complex frequency dependent dielectric constant and conductivity, respectively. Especially the frequency dependence of $\epsilon(\omega)$ is important because it can largely vary for different substrate materials.

For our modeling we simplify by using $\epsilon(\omega)$ for Sapphire. The resistivity of the stack is approximated by using $\sigma(\omega)$ for Pt. Both frequency dependencies are taken from literature \cite{1985querry, Rakic98}. To obtain $k$ we first calculate the transfer function $F_T$ from the experiment by dividing the spectral intensity for one type of nanoSTE |S$\mathrm{_{nSTE}(\omega)}$| by the spectral intensity of the reference emitter |S$\mathrm{_{ref}(\omega)}$|. Starting from this reference spectrum allows us to neglect all spectral modifications that occur due to the setup and the detection scheme. These are difficult to determine, but they are identical for all measurements.
We then fit $F_T$ with only $k$ as a free parameter and only for a single value of $w$. 
The fitting parameter $k$ for the time constant $RC$ is 120 for a 900\,nm nanoSTE.

By using $k\propto w$ ($k$ $\mathrm{ = 240, 600, 1200}$ and $2400$ for $w$ $\mathrm{ = 1.8\,\mu m, 4.5\,\mu m, 9\,\mu m}$ and $\mathrm{18\,\mu m}$, respectively), we must now be able to reproduce $F_T$ for all other nanoSTE sizes. 
The result is shown in Fig. \ref{fig3}(c).
We find a good agreement with the measurements shown in Fig. 2(b). On the one hand, we observe the high pass characteristics predicted in \cite{chargedynamics}. The cut-off frequency increases with decreasing $w$ because $R$ and thus also $RC$ become smaller. On the other hand, we also reproduce the dips at 15 THz and 24 THz. These appear because  both imaginary and real part of $\epsilon(\omega)$ of sapphire exhibit maxima and/or minima around those frequencies, that reduce $F_T$ at those frequencies. Plotting $F_T$ itself and comparing it to the transfer function calculated from the different measurements in Fig. \ref{fig3}(d) shows even a good quantitative agreement. The comparison of the phase of the transfer function for theory and experiment are shown in the supporting information (Fig. S2). Also shown there are the results of a similar experiment with structures on a glass substrate and the corresponding calculated spectra based on the dielectric properties of glass.

\subsection{Coupling due to the spacing between the nanoSTEs}
To further verify the model, we include the influence of $C_g$ by investigating arrays with identical square size `$w$' but different respective spacing `$s$'. Fig \ref{fig4}(a)-(c) shows structures and the transfer functions. The transfer functions are normalized by the filling factor, because increased spacing also means a much reduced signal intensity. The measurements  show small but noticeable effects in the low frequency part of the spectrum, while the dips are barely modified. While for decreasing structure size the resistance was changing, now the capacitance $C_g$ is slightly decreased with increased spacing, shifting the cut-off frequency to higher values. This also slightly affects the dips at higher frequency but not as much as the low frequency part as is confirmed by the modelling. A quantitative analysis shows that the change of $C_g$ modifies the k-factor by approx. 50 $\%$. In the supporting information, we provide the phase data for these experiments (Fig. S3(a)-(b)).

\subsection{Dependence of width and height}

A final set of experiments only serves to exclude that we are simply dealing with a size effect. By using rectangular emitters and applying the magnetic field either along the short or the long edge of the structures, we can see that it is only the change in $w$ that modifies the time constant while changing $h$ does not influence the emission spectrum. This is predicted by the model and as Fig. \ref{fig4}(d)-(i) shows confirmed by the experiment. In the supporting information we also provide the phase data for these experiments (Fig. S3(c)-(f)).\\

In conclusion we can state that for patterned STE that are smaller than the laser spot size and thus homogeneously illuminated, we see massive size effects on the emission spectrum. Changing the dimension in the direction of the current flow  changes the resistance relevant for the back-flow and thus the $RC$ time constant of the system. This results in a controlled high pass filtering of the THz signal and confirms the theoretical model\cite{chargedynamics} but also in large dips in the spectrum occurring at maxima of the substrates dielectric constant. Based on these results one can now intentionally modify and shape the THz spectrum of STE. For example, by choosing a suitable emitter size one can achieve a well defined lower cut-off frequency for an STE. This can be especially relevant when experiments require only higher THz frequencies but would be adversely affected by lower parts of the spectrum. Or, by selecting a suitable substrate material, it is possible to suppress certain frequencies in the emission band, the degree of suppression determined by the size of the nano emitters.

\begin{acknowledgements}

We acknowledge the German Research Foundation (DFG) for
financial support through the Collaborative Research Center
TRR 227 Ultrafast Spin Dynamics, Project B02. 

\end{acknowledgements}


\providecommand{\noopsort}[1]{}\providecommand{\singleletter}[1]{#1}%

\end{document}


\preprint{APS/123-QED}

\title{Shaping THz emission spectra by using \\ sub-wavelength nanopatterned spintronic THz emitters}

\author{Bikash Das-Mohapatra}
\affiliation{Institut f\"{u}r Physik, Martin-Luther-Universit\"{a}t Halle-Wittenberg, Von-Danckelmann-Platz 3, 06120 Halle, Germany}

\author{Reza Rouzegar}
\affiliation{Department of Physical Chemistry, Fritz Haber Institute, Faradayweg 4-6, 14195 Berlin, Germany}

\author{Evangelos Th. Papaioannou}
\affiliation{Institut f\"{u}r Physik, Martin-Luther-Universit\"{a}t Halle-Wittenberg, Von-Danckelmann-Platz 3, 06120 Halle, Germany}
\affiliation{Department of Physics, Aristotle University of Thessaloniki, 54124 Thessaloniki, Greece}
 
\author{Tobias Kampfrath}
\affiliation{Department of Physical Chemistry, Fritz Haber Institute, Faradayweg 4-6, 14195 Berlin, Germany}

\author{Georg Schmidt}
\email{georg.schmidt@physik.uni-halle.de}
\affiliation{Institut f\"{u}r Physik, Martin-Luther-Universit\"{a}t Halle-Wittenberg, Von-Danckelmann-Platz 3, 06120 Halle, Germany}
\affiliation{Interdisziplin\"{a}res Zentrum f\"{u}r Materialwissenschaften,
Martin-Luther-Universit\"{a}t Halle-Wittenberg, Heinrich-Damerow-Strasse 4, 06120 Halle, Germany
}

\date{\today}

\maketitle

\begin{figure}[h]
\renewcommand{\thefigure}{S\arabic{figure}}
{\includegraphics{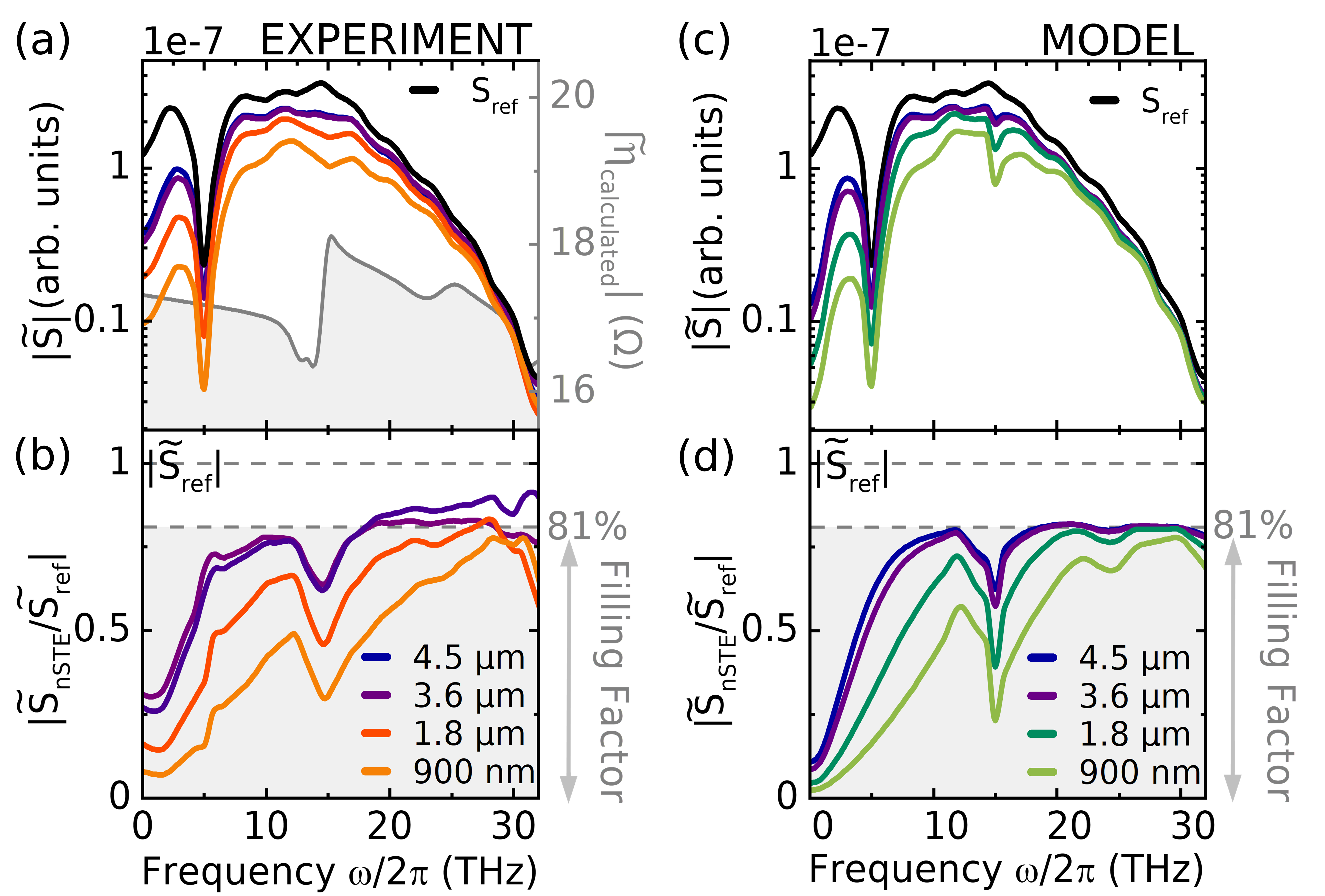}}

   \caption{Experimental results and calculated spectra for nano STE on glass.(a)Spectra $|S(\omega)|$ obtained for different square sizes by FFT of the Electro-Optic signal after applying a von Hann window. The grey line shows the calculated characteristic wave impedance $|\tilde{\eta}(\omega)|$ for the glass substrate. (b) Ratio between spectra for nanoSTEs and Thin Film STE to show the transfer function. (c)Spectra  $|S(\omega)|$ for different sized nanoSTEs calculated applying the circuit Model on the thin film STE reference spectrum (on glass). (d) The calculated transfer function for the nano STE on glass nicely reproduces the measured data from Fig. \ref{figS1}(b)
}
\label{figS1}
\end{figure}

To further confirm the validity of our model we have done similar experiments with nanoSTE on glass substrates which is shown in Fig. \ref{figS1}. The frequency dependence of the dielectric constant of glass is different than for the sapphire substrate used in the main manuscript. Glass also has a strong non-monotonicity or a peak at approx. 15 THz but no distinct changes above 20 THz. Also for this substrate the high pass filter characteristics at low frequencies are observed. In addition we see a dip at 15 THz that increases for smaller structures as we expect from the model. The characteristic dip observed for sapphire substrate at 24 THz, however, is absent, as would be expected from the dielectric constant. Using the dielectric properties of glass, our model nicely reproduces these features also for this different substrate.

\begin{figure}[!ht]
\renewcommand{\thefigure}{S\arabic{figure}}
{\includegraphics[width=0.41\columnwidth]{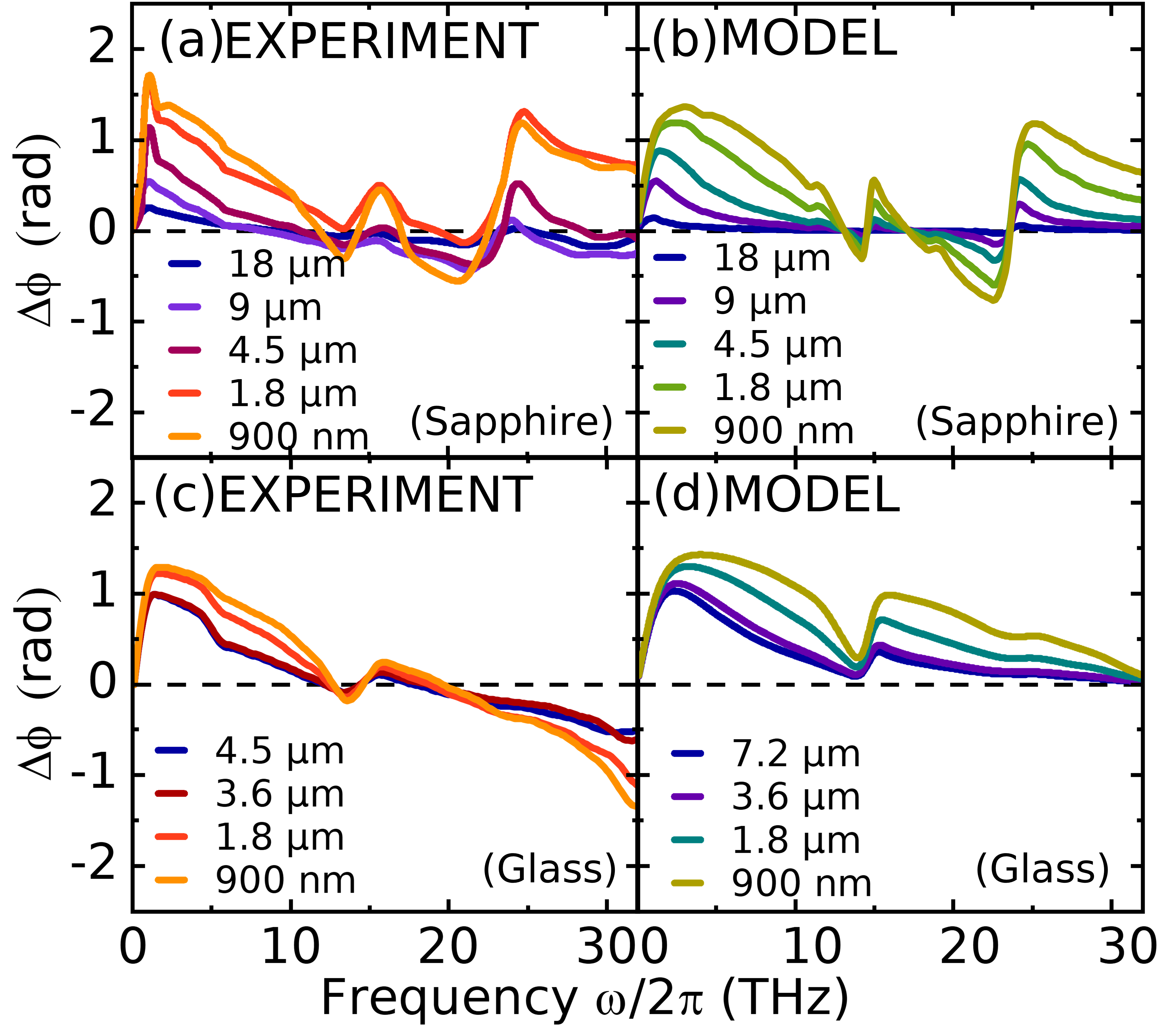}}

   \caption{Comparison of the phase of the transfer function as determined from (a) experiments on STE on sapphire substrates and calculated from the high pass filter function with the appropriate dielectric function for sapphire (b). (c) and (d) show the same analysis for the STE on glass.
}
\vspace{-1em}
\label{figS2}
\end{figure}

Corresponding to the magnitude of the transfer function, we have also analysed the phase in theory and experiment. As shown in the main manuscript, the magnitude or the respective transfer function can be obtained by applying the filter function to the signal measured for a large area emitter. For the phase the situation is more complex, because as described before\cite{chargedynamics}, even in a large area emitter charging phenomena occur. While in the relevant frequency range this barely influences the magnitude, the phase can already change considerably resulting in the necessity to apply a k-factor also for the large area STE. As Fig. \ref{figS2} shows, applying the high pass filter function using $k$ $\mathrm{ = 6000}$ for $\mathrm{STE_{ref}}$ with the appropriate dielectric function of the sapphire substrate to the measurement of a large area STE nicely reproduces the phase as it changes for the differenty sized nano emitters. 

\begin{figure}[!ht]
\renewcommand{\thefigure}{S\arabic{figure}}
{\includegraphics[width=0.46\columnwidth]{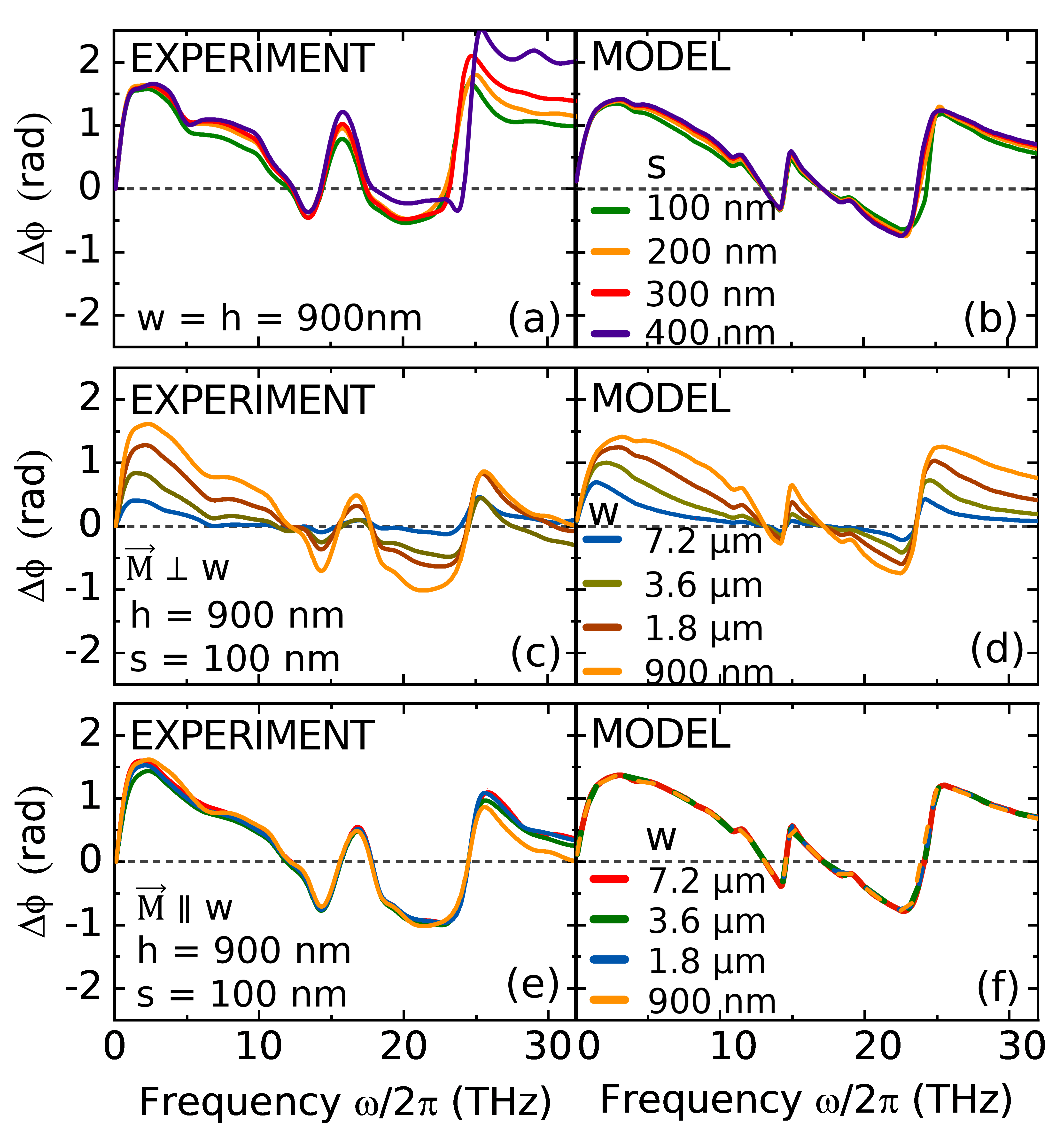}}
\vspace{-1.5em}
   \caption{Comparison of the phase of the transfer function as determined from experiments on STE on sapphire substrates and calculated from the high pass filter function with the appropriate dielectric function for sapphire for different spacing in square nanoSTEs ((a) and (b) respectively), for rectangular nanoSTEs when $\vec M \perp w$ ((c) and (d) respectively) and $\vec M \parallel w$ ((e) and (f) respectively).
}
\vspace{-2em}
\label{figS3}
\end{figure}

Finally we have extracted the phase data from the measurements of emitters with different spacing and the rectangular emitters. In Fig. \ref{figS3} this data is shown and compared to that obtained by the analytical model. We find that both are in good agreement.

\providecommand{\noopsort}[1]{}\providecommand{\singleletter}[1]{#1}%